\documentstyle[12pt]{article}
\def\href#1#2{{#2}}
\begin{document}
\begin{titlepage}
\begin{flushright}
quant-ph/9806023 \\
TS-TH-98-1 \\
June 1998
\end{flushright}
\vspace{12pt}
\begin{center}
\LARGE
Bohm Confirmed \\
by NonRelativistic Quark Model \\
\vspace{24pt}
\normalsize
Frank D. (Tony) Smith, Jr. \\
\vspace{6pt}
\footnotesize
e-mail: fsmith@pegasus.cau.edu \\
and tsmith@innerx.net  \\
P. O. Box 370, Cartersville, GA 30120 USA \\
WWW URLs http://galaxy.cau.edu/tsmith/TShome.html \\
and http://www.innerx.net/personal/tsmith/TShome.html \\
\end{center}
\normalsize
\vspace{24pt}

\begin{abstract}

The effectiveness of the NonRelativistic Quark Model of hadrons  \\
can be explained by Bohm's quantum theory applied to a fermion \\
confined in a box, in which the fermion is at rest because its kinetic \\
energy is transformed into PSI-field potential energy. \\
Since that aspect of Bohm's quantum theory is not a property of \\
most other formulations of quantum theory, \\
the effectiveness of the NonRelativistic Quark Model confirms \\
Bohm's quantum theory as opposed to those others.  \\

\end{abstract}
\normalsize
\end{titlepage}
\newpage
\setcounter{footnote}{0}
\setcounter{equation}{0}

\section{Bohm and NonRelativistic Quark Model}

The effectiveness of the NonRelativistic Quark Model of hadrons
can be explained by Bohm's quantum theory applied to a fermion
confined in a box, in which the fermion is at rest because
its kinetic energy is transformed into potential energy of
interaction with the $\Psi$-field.
Since that aspect of Bohm's quantum theory is not a property of
most other formulations of quantum theory,
the effectiveness of the NonRelativistic Quark Model confirms
Bohm's quantum theory as opposed to those others.

\vspace{12pt}

Bohm's Hidden Variable papers I and II, were published in
the Physical Review in 1952 \cite{DB}, well before QCD was known.

\vspace{12pt}

Although Bohm's theory does not explain why quarks
are confined by QCD inside hadrons,
if you assume that quark confinement by QCD acts to
contain each valence quark fermion within the
impenetrable and perfectly reflecting boundary of the hadron,
then Bohm's Theory explains why the NonRelativistic Quark Model
of light-quark hadrons works so well.

\vspace{12pt}

Consider paper II, section 5, which is reprinted at page 387
of \cite{WZ}:

\vspace{12pt}

"A more striking illustration ... is afforded by the
problem of a "free" particle contained between
two impenetrable and perfectly reflecting walls,
separated by a distance L.
For this case, the spatial part of the $\Psi$-field is

$$
\Psi = sin(2 \pi n x / L)
$$

where n is an integer and the energy of the electron is

$$
E = ( 1 / 2 m ) ( n h / L )^2
$$

Because the $\Psi$-field is real, we deduce that the particle
is at rest.

\vspace{12pt}

"Now, at first sight, it may seem puzzling that
a particle having a high energy should be at rest
in the empty space between two walls.
Let us recall, however, that the space is not really empty,
but contains an objectively real $\Psi$-field that can act
on the particle. Such an action is analogous to
(but of course not identical with)
the action of an electromagnetic field,
which could create non-uniform motion of the particle
in this apparently "empty" enclosure.
We observe that in our problem, the $\Psi$-field is able
to bring the particle to rest and to transform
the entire kinetic energy into potential energy of interaction
with the $\Psi$-field. To prove this, we evaluate
the "quantum-mechanical potential" for this $\Psi$-field

$$
U = \frac{ - {h}^{2}}{2m} \frac{{\bigtriangledown}^{2} R}{R}
= \frac{ - {h}^{2}}{2m} \frac{{\bigtriangledown}^{2} \Psi }{\Psi}
= \frac{1}{2m} {\frac{nh}{L}}^{2}
$$

and
note that it is precisely equal to the total energy, $E$."

\vspace{12pt}

\vspace{12pt}

If you apply the Bohm result for an electron confined between
two walls to the QCD picture of a light quark confined
within a hadron, then the Bohm result may explain the effectiveness
of the NonRelativistic Quark Model of light-quark hadrons.

\vspace{12pt}

Conversely:
the effectiveness of the NonRelativistic Quark Model of light-quark
hadrons may be considered to be experimental support for  Bohm's theory.

\vspace{12pt}

The NonRelativistic Quark Model of light-quark hadrons is described
in many textbooks, including \cite{MG} by Mike Guidry, who,
at page 381 (see also page 311) says:

\vspace{12pt}

"By uncertainty principle agruments the momentum of
a quark confined to the radius of one fermi
is [about 200 MeV] ... For u or d quarks [the currrent mass is
about 10 MeV or less and the constituent mass is about 300 MeV] ...
and a nonrelativistic approximation is questionable ... relativity
effects should be significant. Nevertheless, nonrelativistic models
of quark structure for hadrons have been found to work
surprisingly well, even for light hadrons. ..."

\end{document}